\documentclass[9pt,twocolumn,twoside]{osajnl}
\journal{ol}
\setboolean{shortarticle}{true}

\newcommand\unit[2]{\ensuremath{#1~\mathrm{{#2}}}}

\title{Crossed optical cavities with large mode diameters}

\author[1,2]{André Heinz}
\author[1,2]{Jan Trautmann}
\author[1,2]{Neven Šantić}
\author[1,2]{Annie Jihyun Park}
\author[1,2,3]{Immanuel Bloch}
\author[1,2,*]{Sebastian Blatt}

\affil[1]{
  Max-Planck-Institut f{\"u}r Quantenoptik,
  Hans-Kopfermann-Stra{\ss}e 1,
  85748 Garching, Germany}
 \affil[2]{
  Munich Center for Quantum Science and Technology,
  80799 M{\"u}nchen, Germany}
\affil[3]{
  Fakult{\"a}t f{\"u}r Physik,
  Ludwig-Maximilians-Universit{\"a}t M{\"u}nchen,
  80799 M{\"u}nchen, Germany}

\affil[*]{Corresponding author: sebastian.blatt@mpq.mpg.de}

\dates{Compiled \today}

\doi{\url{http://dx.doi.org/10.1364/XX.XX.XXXXXX}}

\begin{abstract}
  We report on a compact, ultrahigh-vacuum compatible optical assembly to create large-scale, two-dimensional optical lattices for use in experiments with ultracold atoms.
  The assembly consists of an octagon-shaped spacer made from ultra-low-expansion glass, to which we optically contact four fused-silica cavity mirrors, making it highly mechanically and thermally stable.
  The mirror surfaces are nearly plane-parallel which allows us to create two perpendicular cavity modes with diameters $\sim$\unit{1}{mm}.
  Such large mode diameters are desirable to increase the optical lattice homogeneity, but lead to strong angular sensitivities of the coplanarity between the two cavity modes.
  We demonstrate a procedure to precisely position each mirror substrate that achieves a deviation from coplanarity of $d = 1(5)\,\mu m$.
  Creating large optical lattices at arbitrary visible and near infrared wavelengths requires significant power enhancements to overcome limitations in the available laser power.
  The cavity mirrors have a customized low-loss mirror coating that enhances the power at a set of relevant wavelengths from the visible to the near infrared by up to three orders of magnitude.
\end{abstract}

\setboolean{displaycopyright}{true}

\begin{document}

\maketitle

Fabry-P\'erot resonators~\cite{kogelnik66,vaughan89} for visible and near-infrared light are fundamental tools of laser science and atomic, molecular, and optical physics.
They are used to provide feedback around a laser gain medium, as mechanical length references for optical frequency standards~\cite{robinson19}, as strongly-coupled interfaces between quantum emitters and light~\cite{dutra05}, to sense tiny forces  induced by gravitational waves~\cite{aasi15}, to enhance the spectroscopic signals from optically thin samples~\cite{berden10}, and to create deep optical traps for ultracold atoms~\cite{mosk01}.
The most important cavity design parameter for each use case is the finesse $\mathcal{F}$, which determines the cavity's power enhancement and its frequency selectivity.
Decades of technological improvements have enabled $\mathcal{F}$ up to $5\times 10^{5}$~\cite{robinson19,harry12}.
\begin{figure}[ht]
\centering
\includegraphics[width=\linewidth]{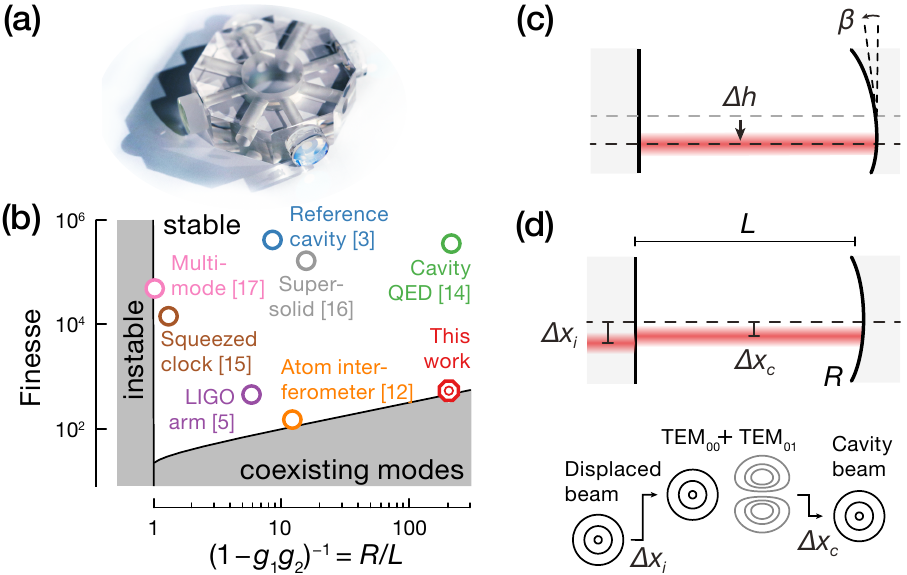}
\caption{(a) Photograph of the cavity assembly. (b) Different applications of Fabry-P\'erot resonators can be classified according to their demands on finesse and stability. For plano-concave resonators, the stability criterion $0 \le g_1 g_2 \le 1$ can be recast in terms of the ratio of mirror curvature radius $R$ to cavity length $L$. As $R/L$ increases, the cavity enters the near plane-parallel regime, where two systematic effects are important. (c) A relative wedge angle $\beta$ between both mirrors leads to a transverse shift of the optical axis $\Delta h$, which becomes important when crossing two independent resonators. (d) A displaced input beam leads to a displaced cavity beam, an effect that is suppressed with increasing finesse as the cavity linewidth narrows compared to the transverse mode splitting.}
\label{fig:design}
\end{figure}
The second important design parameter is the diameter of the optical mode in the cavity, given by the radius of curvature $R_1$ and $R_2$ of the mirrors and the mirror spacing $L$.
Stable Hermite-Gaussian eigenmodes form between the mirrors when the stability parameter product $0 \leq g_1 g_2 \leq 1$, where $g_i = 1 - L/R_i$~\cite{kogelnik66}.
Stable cavities are desirable because they provide clearly distinguishable eigenmodes that allow working with a given laser beam geometry and a well-defined resonance frequency.
However, it can be advantageous to work with nearly instable cavities: even though they are sensitive to mechanical and thermal imperfections, large mode diameters on short cavity lengths can be achieved.
Increasing the mode diameter is a long-standing problem in experiments with ultracold atoms trapped in far-off-resonant optical lattices, formed by the antinodes of interfering laser beams.
Stable trapping requires high laser intensities to achieve deep traps and low laser noise to prevent heating the atoms~\cite{blatt15}.
Depending on the wavelength required for the optical lattice, the available laser power sets a maximal beam diameter which leads to a hard technical limit on how many atoms can be used.
Large mode diameters are not only advantageous for trapping higher numbers of atoms and therefore scaling to larger system sizes, but also for mitigating inhomogeneous trapping effects.
These effects often crucially limit the performance of quantum simulators~\cite{gross17}, where e.g. correlation lengths can be cut off by the inhomogenous trapping environment.
For this reason, applications such as quantum simulators~\cite{gross17}, optical frequency standards~\cite{origlia18}, and atomic sensors~\cite{hamilton15}, can benefit from power enhancement cavities with large mode diameters.

In this paper, we demonstrate and characterize cavities with mode diameters $\sim$\unit{1}{mm} that operate in the near plane-parallel regime with $g_1 g_2 \simeq 0.995$.
For this purpose, we use one flat and one curved mirror with $R = \unit{10.2}{m}$, spaced by $L=\unit{50}{mm}$.
We present an optical assembly which has no adjustable parts and contains two such cavities whose fundamental modes cross at right angles.
The assembly, shown in Fig.~\ref{fig:design}(a), consists of two pairs of cavity mirrors, optically contacted~\cite{tong98} to an octagon-shaped spacer such that the fundamental modes of both cavities are coplanar, are long-term stable, and do not require alignment.
When laser light is coupled into each resonator, a two-dimensional optical lattice forms in the region where both modes cross.
The spacer is made from ultra-low-expansion glass and the mirror coatings are ion-beam-sputtered onto fused silica substrates, which ensures vacuum compatibility and a high thermal stability.
We choose a moderate finesse $\mathcal{F} = 3\times 10^2 - 5 \times 10^3$ to enable power enhancements $\Lambda = 10^2 - 10^3$ for multiple wavelengths $\lambda$.
As shown in Fig. \ref{fig:design}(b) our choice of a moderate finesse and near plane-parallel mirror geometry distinguishes us from other cavity applications, such as in cavity quantum electrodynamics~\cite{brennecke07}, spin squeezing~\cite{pedrozo20}, reference cavities~\cite{robinson19}, cavity interferometers~\cite{hamilton15}, gravitational wave detectors~\cite{aasi15}, or near-resonant optical lattices to study self-organization~\cite{leonard17,vaida18}.
In this cavity regime, the coplanarity between the two perpendicular cavity modes becomes sensitive to two systematic effects.

First, the optical axis of each resonator is shifted when both cavity mirrors are tilted with respect to each other by a relative angle $\beta$, illustrated in Fig.~\ref{fig:design}(c).
This shift does not break cylindrical symmetry if the curved mirror surface remains spherical around the point of incidence.
The relative shift of each optical axis dominates over all other systematic effects in the near plane-parallel regime. Each optical axis shifts according to
\begin{equation}
  \label{eq:oa_shift}
  \Delta h = R \sin\beta,
\end{equation}
which is completely independent of $L$~\cite{anderson84}.
Our goal of achieving large $1/e^2$ mode diameters
\begin{equation}
  \label{eq:waist}
  2w = 2\sqrt{\frac{\lambda}{\pi}}[L R (1 - L/R)]^{1/4} \propto R^{1/4},
\end{equation}
thus must be balanced against the difficulty of polishing glass spacers with extreme requirements on surface parallelism.

Second, the mode separation of higher-order transverse electromagnetic field (TEM) modes becomes comparable to the linewidth of the cavity resonances, such that we approach a regime of ``coexisting modes,'' as illustrated in Fig. \ref{fig:design}(d).
Here, the optical axis of each cavity is unaffected, but a misalignment $\Delta x_i$ of the input beam leads to a shift $\Delta x_c \equiv \varepsilon \Delta x_i$ of the light circulating in the cavity.
Even when the input beam is frequency-stabilized to the fundamental TEM$_{00}$ resonance, a fraction of the input power given by the ratio of transverse mode splitting to the cavity linewidth is coupled into the (say) TEM$_{01}$ mode~\cite{anderson84}.
If we require a misalignment suppression factor $\varepsilon$, we find a lower bound on the required cavity finesse (see Supplement 1)
\begin{equation}
  \label{eq:regime}
\mathcal{F} \geq \frac{\pi}{2} \frac{\sqrt{1/\varepsilon^2 - 1}}{\arccos\sqrt{g_1g_2}},
\end{equation}
which we show in Fig.~\ref{fig:design}(b) for $\varepsilon = 5\%$ as a new bound in the stability diagram.
For this $\varepsilon$ and a mode-matching to the TEM$_{00}$ mode of 99\%, our cavity design suppresses $\Delta x_c < 1\,\mu\mathrm{m}$.
From the perspective of ray optics we can understand this effect as the minimum number of cavity round trips required to force the average ray offset to lie on the optical axis within a certain margin of error.

Our cavity spacer surfaces have a parallelism $\le\unit{1}{arcsec}$ to suppress $\Delta h$ to $\sim$$49\,\mu\mathrm{m}$.
The octagon-shaped spacer is \unit{50}{mm} wide and \unit{15}{mm} high, which allows attaching mirror substrates with \unit{12.7}{mm} diameters to its sides.
These sides have a \unit{4}{mm} wide bore to allow the formation of cavity modes.
Further optical access is enabled by \unit{5}{mm}-wide bores in the remaining sides of the octagon, and a \unit{20}{mm}-wide bore in its center allows imaging the optical lattices with a microscope objective.

\begin{figure}[t]
\centering
\includegraphics[width=\linewidth]{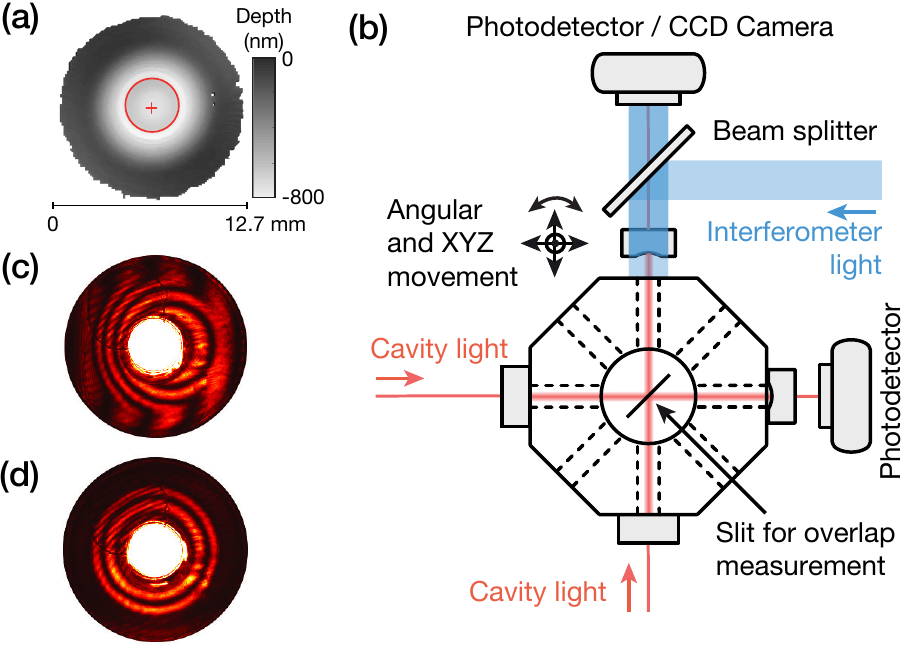}
\caption{(a) Surface profile of a curved mirror substrate with a polished ring. The cross denotes the deepest point of the curved area, and the coating is applied within the circled area. (b) Final step of the building process: after all mirrors except one curved mirror substrate are attached, the optical axis of the first cavity mode is measured. The position of the second curved mirror then fixes the optical axis of the second resonator and determines the coplanarity between both resonators. An interferometer and a transmission measurement performed with a slit are used to make the optical axes coplanar. Panels (c) and (d) show typical interferograms to determine the alignment of the polished ring with respect to the spacer top surface.}
\label{fig:assembly}
\end{figure}

Since the curved cavity mirrors are attached by optical contacting, they require a flat ring polished onto the curved substrate.
For these substrates, the ring is a flat region around a shallow dimple that is less than $1\,\mu\mathrm{m}$ deep, as shown in the surface profile in Fig.~\ref{fig:assembly}(a).
A ring with an inhomogeneous radial thickness would result in an effective tilt $\beta$ of the curved region, which would lead to additional optical axis shifts as discussed above.
For our mirror substrates, this tilt (or wedge error) is smaller than \unit{2}{arcsec}.
The combination of shallow mirror surface, small wedge error, and superpolished surface specifications places extreme requirements on the fabrication of the curved mirror substrates.
An ion-beam-sputtered high-reflectivity coating was applied to the mirror substrates after polishing.
The multilayered coating is $5\,\mu\mathrm{m}$ thick and thus protrudes from the dimple in the substrate.
The mirror coating was applied only to the central \unit{3.5}{mm} of the substrate, such that the masked coating fits the spacer's \unit{4}{mm} diameter bores and leaves enough room to position the mirror.
Furthermore, the uncoated ring allows a high bonding strength when the mirror is attached to the spacer.
The substrates are \unit{6.35}{mm} thick, which reduces deformation of the substrate due to mechanical stress induced by the coating.

Since the final assembly has no adjustable parts, we need to ensure that the two perpendicular cavity modes cross in the center of the assembly by precisely positioning the cavity mirrors.
We optically contact the mirrors to the spacer in a clean-room environment~\cite{heinz20} after having cleaned all glass parts using methods developed for semiconductor fabrication~\cite{tong98}.
The spacer together with a mirror is mounted on a stage that fixes the spacer while the mirror to be contacted can be tilted and translated with respect to the surface, as illustrated in Fig.~\ref{fig:assembly}(b).
We use an imaging system and a camera to position the coated region of the mirror with respect to the spacer bore.
The mirror is then pressed onto the spacer to initiate the optical bond~\cite{tong98}.
With this setup, we measured a lateral positioning uncertainty between initial and contacted mirror position of $30(5)\,\mu\mathrm{m}$.
Using these methods, we attach the first mirror pair, which fixes the orientation and position of the first cavity mode.
After that, the second flat mirror is attached to the spacer, and consequently the final curved mirror determines the optical axis of the second cavity and thereby the coplanarity between the two cavity modes.

Before we contact the final mirror, we measure the coplanarity of both modes in situ.
For this purpose, we recover an interferogram between the lower mirror surface and the spacer top surface, as shown  in Fig.~\ref{fig:assembly}(b).
The interferogram allows us to monitor the relative angle $\beta$ between the polished ring and the spacer surface.
A relative tilt leads to straight fringes in the outer regions of the interferogram and a minimal $\beta$ results in a mostly homogeneous fringe as depicted in Figs.~\ref{fig:assembly}(c) and (d).
The angular and lateral alignment is limited by optical resolution: a residual fifth of a fringe corresponds to a residual $\beta = \unit{1}{arcsec}$.

\begin{figure}[t]
\centering
\includegraphics[width=\linewidth]{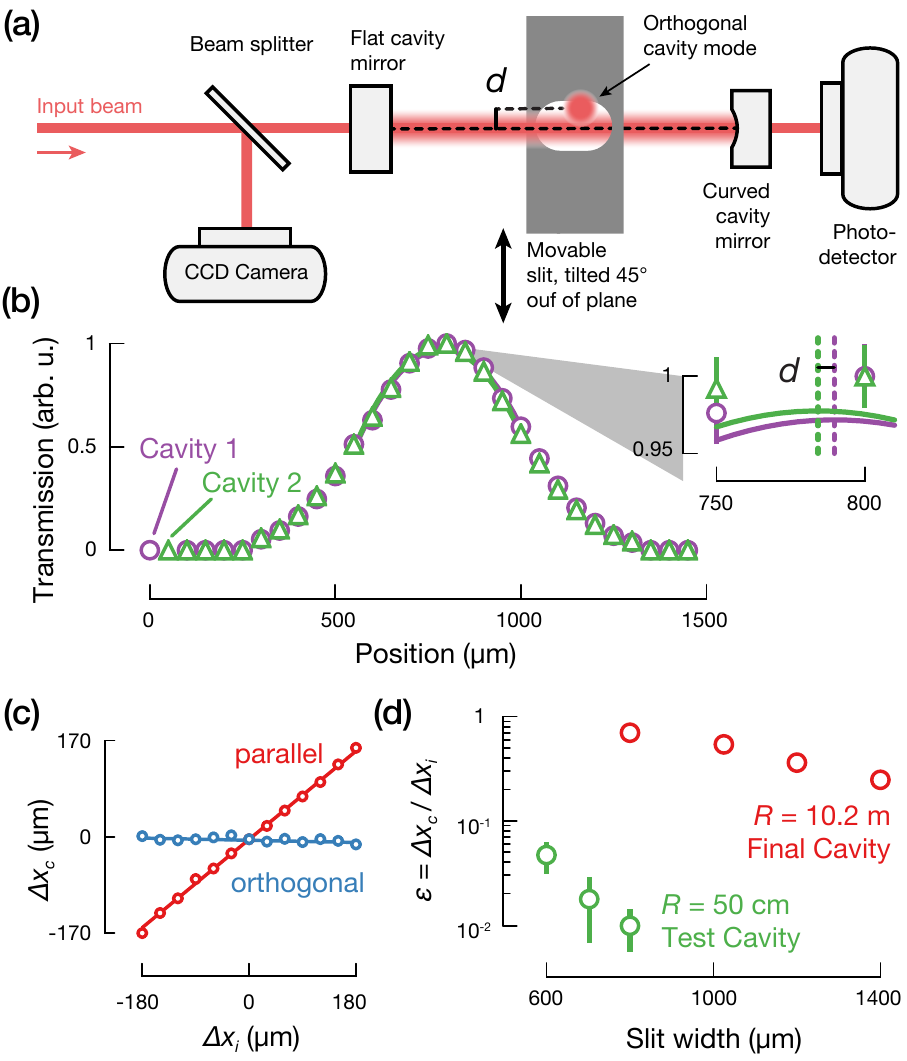}
\caption{(a) Principle of the coplanarity measurement. A laser is frequency-scanned over the cavity free-spectral-range such that two TEM$_{00}$ mode transmission peaks can be observed on photodetectors (PD). A slit is moved vertically across both cavity modes which are separated by a relative displacement $d$. The camera in front of the cavity relates the input beam position to the measured maximum to investigate mode-shifting systematics. (b) Example of a coplanarity measurement. (c) Example measurement of the mode shift suppression factor $\varepsilon$ for a $800\,\mu\mathrm{m}$-wide slit. The red (blue) curve shows the effect of varying $\Delta x_i$ parallel (perpendicular) to the slit measurement direction.
(d) Smaller slit widths reduce the effective finesse of the cavity and increase $\varepsilon$, which can become the dominant systematic effect in the near plane-parallel regime.}
\label{fig:overlap}
\end{figure}

After prealigning the final mirror in this way, we measure the mode coplanarity with a slit that clips parts of both cavity modes.
For the coplanarity measurement, the input beam is frequency-modulated over a free-spectral range, while the slit is positioned using a translation stage.
During this process, we measure the transmission of the TEM$_{00}$ mode for both cavities with separate photodetectors, as shown in Fig.~\ref{fig:overlap}(a).
To understand the transmission signals, we can think of cutting the cavity mode with one side of the slit as cutting a Gaussian beam with a knife edge which would yield an error function.
Because the slit has two edges, the cavity transmission signal varies similar to a sum of two error functions distributed symmetrically around the mode center.
In contrast to a traditional knife-edge measurement, the cavity mode can diffract and circulate around a partial obstruction such that 50\% transmission does not indicate the edge being at the center of the beam profile.
For this reason, we choose a wide enough slit to observe a clear transmission peak (comparable to the $1/e^2$ mode diameter), while still obstructing the wings of the mode enough to result in a clear maximum in the TEM$_{00}$ transmission.
In Fig.~\ref{fig:overlap}(b), we fit a parabola to all points above 50\% of the maximum peak transmission to determine the mode center.
Fitting both mode centers yields the displacement $d$ of both modes perpendicular to the plane spanned by both optical axes.
Once we have minimized $d$, we contact the final cavity mirror to the spacer.
If necessary, we detach the mirror using a razor blade and repeat the contacting procedure.

The slit-based measurement of $d$ is affected by input beam misalignment according to Fig.~\ref{fig:design}(d).
The slit artificially reduces $\mathcal{F}$ by obstructing the wings of each mode and we enter the regime of coexisting modes.
For a slit width of $800\,\mu\mathrm{m}$ and a mode with $w =396\,\mu\mathrm{m}$ at \unit{689}{nm}, we find $\mathcal{F} = 533(19)$ without and $167(38)$ with the slit, respectively.
To determine the significance of input beam misalignment, we vary the input beam position $\Delta x_i$ from the point of maximum mode matching to TEM$_{00}$, and determine $\Delta x_c$ by finding the transmission maximum with respect to the slit position.
In Fig.~\ref{fig:overlap}(c), we show $\Delta x_c$ versus $\Delta x_i$, from which we determine the mode displacement suppression factor $\varepsilon$ with a linear fit.
In the presence of the slit, we measure $\varepsilon = 0.68(1)$ and $-0.02(1)$, when $\Delta x_i$ is altered perpendicular and parallel to the slit.
In Fig.~\ref{fig:overlap}(d), we show how $\varepsilon$ decreases with slit width and verify that $\varepsilon$ decreases dramatically for mirrors with smaller $R$ for the same mirror coating.

To measure the coplanarity $d$ faithfully, we optimized the mode-matching to TEM$_{00}$ to $\sim$$99\%$ multiple times and measured a standard deviation of $\Delta x_i$ of $\sigma_i = 14\,\mu\mathrm{m}$ using the camera shown in Fig.~\ref{fig:overlap}(a).
For the final assembly, we determine the coplanarity using 8 independent measurements of $d$, by rotating and flipping the spacer in its mount with respect to the slit.
After weighted averaging, we find $d = 1(5)\,\mu\mathrm{m}$, where the uncertainty is dominated by $\varepsilon \sigma_i$.
We did not observe any changes in $d$ after repeatedly annealing the assembly at \unit{190}{^\circ{}C} under vacuum. As expected~\cite{tong98}, this procedure strengthens the optical bonds between mirrors and spacer.

\begin{figure}[t]
\centering
\includegraphics[width=\linewidth]{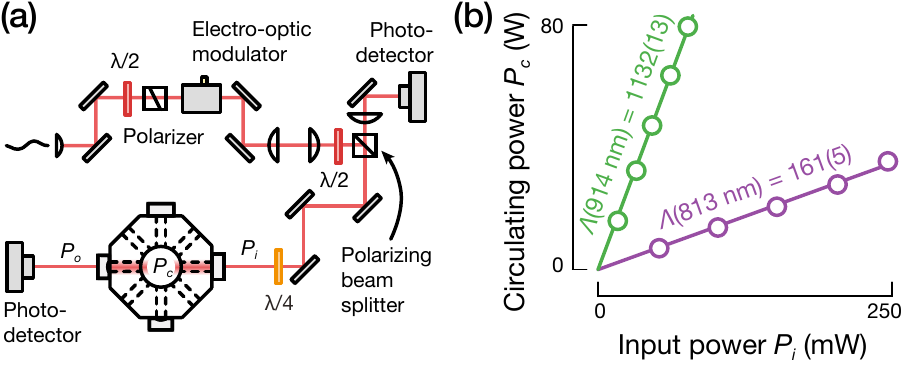}
\caption{(a) Setup to measure the power buildup coefficient $\Lambda$ of a cavity. (b) Examples of a power buildup measurements.}
\label{fig:buildup}
\end{figure}

We characterize the power enhancement factor $\Lambda = P_c / P_i$ for each wavelength of interest by estimating the circulating power $P_c$ for input (output) power $P_i$ ($P_o$) using the setup shown in Fig.~\ref{fig:buildup}(a).
Our method can be applied when the finished assembly has been placed inside a vacuum chamber, and it does not assume that the cavity is lossless~\cite{heinz20}.
We first measure $\mathcal{F}$ from a transmission spectrum, and with cavity ring-down spectroscopy~\cite{berden10}.
We then extract the amplitude reflection coefficient of a single mirror $r = \sqrt{\pi^2 / (4\mathcal{F}^2) + 1} -\pi/(2\mathcal{F})$.
The circulating power is $P_c = \sqrt{P_i P_o}/(1-r^2)$, where $P_o$ is measured while the laser is locked to a cavity arm via a Pound-Drever-Hall lock~\cite{heinz20}.
By varying $P_i$, we obtain the data shown in Fig.~\ref{fig:buildup}(b).
We find no evidence of nonlinear effects up to circulating powers of \unit{80}{W}, and measure $\Lambda(\unit{689}{nm}) = 147(5)$, $\Lambda(\unit{813}{nm}) = 161(5)$, and $\Lambda(\unit{914}{nm}) = 1132(13)$.
  Compared to applications that require high-finesse cavities~\cite{robinson19}, optical lattices benefit from cavities with moderate $\mathcal{F}$.
  First, dynamic control over the lattice depth is limited by the cavity lifetime $\mathcal{F}/(2\pi\nu_\mathrm{FSR})$, where $\nu_\mathrm{FSR} \simeq \unit{3}{GHz}$ is the free spectral range.
  Second, an increased $\mathcal{F}$ results in narrower cavity resonances, which makes the trapped atoms suspectible to heating~\cite{blatt15} by converting laser frequency noise to lattice amplitude noise.

In conclusion, we have demonstrated an optical assembly containing two perpendicular optical resonators with well-crossed optical axes.
  Each resonator supports a fundamental mode with a $1/e^2$ mode diameter of $(910~\mu\mathrm{m}) \times \sqrt{\lambda / (\unit{914}{nm})}$, which results in an order-of-magnitude increase in available lattice sites compared to state-of-the-art free-space~\cite{blatt15} and ring-cavity-based~\cite{cai20}, two-dimensional optical lattices.
This compact optical assembly provides an optimal crossing of two separate laser beams, is compatible with operation in ultrahigh vacuum, has excellent mechanical and thermal stability, and enhances the optical power for multiple optical wavelengths of interest.
By coupling light into the optical resonators, we can create large and stable optical lattices which benefit quantum simulators, atomic sensors, and optical lattice clocks by providing a large number of homogeneous microtraps for ultracold atoms.
This increase in available lattice sites allows using more atoms to directly improve the signal-to-noise ratio in all these applications.
The assembly contains no moving parts and no chemical adhesives, while the monolithic and symmetric design makes the optical lattice alignment insensitive to accelerations.
This insensitivity also makes our assembly an ideal candidate for integration into transportable sensors and clocks for geodesy~\cite{mehlstaubler18}, or for space missions~\cite{origlia18}.
Our methods may find use in other practical applications such as compact monolithic laser gyroscopes~\cite{ezekiel77,schreiber13}, because they shift the complexity of cavity alignment to the manufacturing stage.

\section*{Funding}

Natural Sciences and Engineering Research Council of Canada (517029); European Research Council (Marie Skłodowska-Curie grant no. 844161); H2020 European Research Council (PASQuanS grant no. 817482, UQUAM grant no. 319278).

\section*{Disclosures}

The authors declare no conflicts of interest. The cavity assembly is subject of an international patent application (PCT/EP2019/066247). The current address of N.~\v{S}. is Institute of Physics, Bijeni\v{c}ka cesta 46, 10000 Zagreb, Croatia.

\bigskip
\noindent See Supplement 1 for supporting content.

%\bibliography{crossed_cavity}
%\bibliographyfullrefs{crossed_cavity}

\end{document}

\maketitle

\section{Derivation of Eqn.~(3) in the main text}
\label{sec:derivation}

Assume that an input beam that would be perfectly mode-matched to the TEM$_{00}$ mode of a cavity with mode waist $w$ is instead displaced by $\Delta x_i \ll w$.
The electric field $E_i$ of this displaced input beam can be written as a superposition of the cavity's TEM$_{00}$ and TEM$_{01}$ mode functions $U_0$ and $U_1$ as~\cite{anderson84}
\begin{equation}
\label{eq:x_in}
E_i(x) = A U_0(x-\Delta x_i) \approx A\left[U_0(x) + \frac{\Delta x_i}{w} U_1(x)\right].
\end{equation}
Assuming that the input beam is frequency-stabilized to the transmission peak of the TEM$_{00}$ mode, we estimate the electric field $E_c$ circulating in the cavity by taking into account that the TEM$_{01}$ mode is coupled off-resonantly, given by the transverse mode separation frequency~\cite{kogelnik66}
\begin{equation}
\label{eq:mode_sep}
\nu_\mathrm{sep} = \frac{\Delta\nu_\mathrm{FSR}}{\pi}\arccos(\sqrt{g_1g_2}),
\end{equation}
where $\nu_\mathrm{FSR} = c / 2L$, and $c$ is the speed of light.
The contribution of the $U_1$ mode function to $E_c$ compared to its contribution to $E_i$ is thus attenuated by $\varepsilon \equiv [1 + 4(\Delta \nu_\mathrm{sep} / \Delta\nu)^2]^{-1/2}$.
For the circulating field, we find
\begin{equation}
\label{eq:x_cav}
E_c(x) \approx A\left[U_0(x) + \frac{\varepsilon \Delta x_i}{w} U_1(x)\right],
\end{equation}
from which we conclude that the displacement of the circulating field is suppressed with respect to the displacement of the input field by $\Delta x_c \equiv \varepsilon\Delta x_i$.
Combining the definition of the finesse $\mathcal{F} = \Delta\nu_\mathrm{FSR}/\Delta\nu$ with Eqn.~\ref{eq:mode_sep}, we find Eqn.~(3) in the main text.

% \bibliography{crossed_cavity}